\documentclass[prl,aps,amssymb,amsfonts,amsmath,showpacs,twocolumn]{revtex4}
\usepackage{graphicx}

\newcommand{\real}{\operatorname{Re}}

\begin{document}

\title{Strong extinction of a laser beam by a single molecule}

\author{I. Gerhardt}
\author{G. Wrigge}
\author{P. Bushev}
\author{G. Zumofen}
\author{R. Pfab}
\author{V. Sandoghdar}\email{vahid.sandoghdar@ethz.ch}

\affiliation{Laboratory of Physical Chemistry, ETH Zurich, CH-8093
Zurich, Switzerland}

\begin{abstract}
We present an experiment where a single molecule strongly affects
the amplitude and phase of a laser field emerging from a
subwavelength aperture. We achieve a visibility of $-6\%$ in
direct and $+10\%$ in cross-polarized detection schemes. Our
analysis shows that a close to full extinction should be possible
using near-field excitation.
\end{abstract}


\date{April 13, 2006}

\maketitle

The strength of the interaction between radiation and matter is
often formulated in terms of a cross section, modelling the object
as a disk of a certain area exposed to the radiation. For an ideal
oriented quantum mechanical two-level system with a transition
wavelength $\lambda$, the absorption cross section amounts to
$\sigma_{abs}=3\lambda^2/2\pi$~\cite{Loudon}. This formulation
suggests that a single emitter would fully extinguish the incident
light if one could only focus it to an area smaller than
$\sigma_{abs}$. In fact, it is possible to confine a laser beam to
an area of about $(\lambda/2 N.A.)^2$, using immersion optics with
a high numerical aperture ($N.A.$) or to confine it to a
subwavelength area using Scanning Near-field Optical Microscopy
(SNOM). However, it turns out that it is a nontrivial task to
explore these ideas experimentally because systems in the gas
phase are not easily compatible with strongly confined laser
fields~\cite{Wineland:87}. Moreover, solid-state emitters suffer
from broad homogeneous linewidths ($\gamma$) at room temperature,
reducing $\sigma_{abs}$ by the factor $\gamma_0/\gamma$ where
$\gamma_0$ is the radiative linewidth~\cite{Loudon}. Considering
these constraints, we have chosen to apply SNOM to excite
molecules at $T=1.4$~K and have succeeded in detecting single
molecules in transmission with a visibility of up to $10\%$, which
we present in this Letter.

Optical detection of single solid-state emitters was pioneered by
applying a double modulation technique in absorption
spectroscopy~\cite{Moerner:89}. In that work molecules doped in an
organic crystal were excited via the narrow zero-phonon lines
(transition $1 \rightarrow 2$, see Fig.~1b) at liquid helium
temperature, and a laser focus of $3~\mu \rm m$ yielded an
absorption effect of about $10^{-4}$. Soon after that it was shown
that the detection signal-to-noise ratio (SNR) would be much more
favorable if one performed fluorescence excitation spectroscopy to
collect the Stokes shifted emission of the molecule (transition $2
\rightarrow 3$) while blocking the excitation beam with
filters~\cite{Orrit:90,SMbook}. Unfortunately, however, this
method sacrifices the information about coherent processes at the
frequency $\nu_{21}$. A few recent works have used the
interference between the coherent radiation of a single emitter
and a part of the excitation beam in reflection to get around this
issue~\cite{Plakhotnik:01,Alen:05}. Here we extend the latter idea
to direct transmission measurements.

\begin{figure}[b!]
\centering
\includegraphics[width=8 cm]{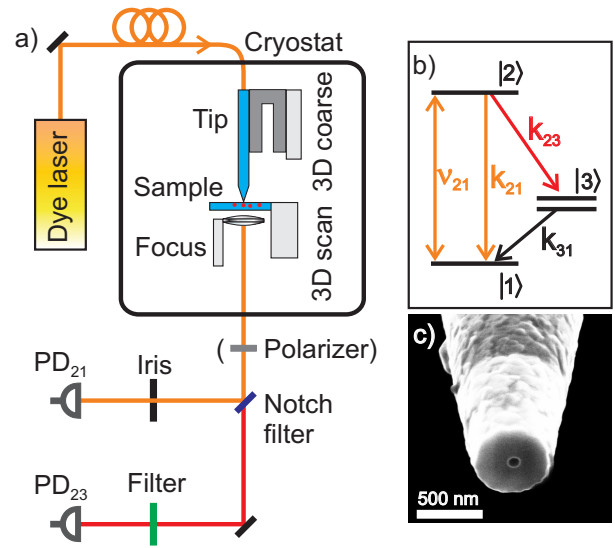}
\caption{(a) Schematics of the experimental setup. The polarizer was
inserted for the data in Fig.~3; for details see text. (b) The
molecular energy level scheme. (c) An electron microscope image of a
typical SNOM probe. }
\end{figure}

Figure 1a shows the schematics of the experimental arrangement
based on a combined cryogenic confocal/near-field microscope
operating at $T=1.4$~K~\cite{Michaelis:00}. Light from a tunable
dye laser ($\lambda \sim 615$~nm, linewidth $\sim$1~MHz) was
coupled into a single mode fiber leading to the SNOM tip in the
cryostat. The SNOM probes were fabricated by evaporating aluminum
onto heat-pulled fiber tips and subsequent focused ion beam
milling of their ends to produce apertures of about 100~nm (see
Fig.~1c). The tip could be positioned in front of the sample using
home-built piezo-electric sliders. The shear force signal using a
quartz tuning fork was used to determine the point of contact
between the tip and the sample to within 10~nm. The sample could
be scanned in 3 dimensions using a commercial piezo-electric
stage. The light transmitted through the tip or emitted by the
molecules was collected by a microscope objective ($N.A.$=0.8)
which could also be translated by a slider. Filters and beam
splitters were used to direct the radiation of the $2 \rightarrow
1$ transition (of the order of $40\%$ of the total emission) to
the avalanche photodiode $PD_{21}$ and the Stokes shifted
fluorescence to $PD_{23}$. In the experiments discussed in this
work, an iris with an opening of about 1~mm selected the central
$10\%$ of the forward emission in the $PD_{21}$ path.

Crystalline films of \emph{p}-terphenyl doped with
dibenzanthanthrene (DBATT) molecules~\cite{Boiron:96} were
fabricated on a glass coverslip by spin coating~\cite{pfab:04}.
After an annealing step, the resulting sample contained large
regions of \emph{p}-terphenyl with typical heights of 50-100~nm,
thin enough to allow near-field studies. By performing room
temperature experiments, we determined the transition dipole
moments of the DBATT molecules to lie at an angle of about $25\pm
5^\circ$ with respect to the tip axis and measured the
fluorescence lifetime of level $\left\vert 2\right\rangle$ to be
$20\pm3$~ns, corresponding to a full width at half-maximum
linewidth of $\gamma_0=8 \pm 1 $~MHz.

\begin{figure}[b!]
\centering
\includegraphics[width=7 cm]{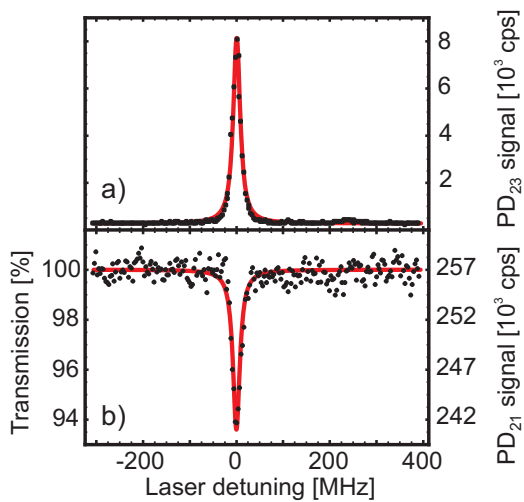}
\caption{Symbols show simultaneously recorded near-field
fluorescence excitation (a) and transmission (b) spectra. The solid
red curves display fits. cps stands for counts per second.}
\end{figure}

In a given experimental run, we first used fluorescence excitation
spectroscopy to identify the zero-phonon transition frequencies of
the molecules located under the tip aperture. Next we adjusted the
tip-sample separation to about 60~nm and optimized the fluorescence
signal $I_{23}$ from the molecule by laterally positioning the
sample in the $x$ and $y$ directions. Fig.~2a shows a near-field
fluorescence excitation spectrum measured on $PD_{23}$. Fig.~2b
displays the signal on $PD_{21}$ recorded simultaneously with the
first spectrum, revealing an impressive drop of $6\%$ in the
incident beam power due to the presence of a single molecule. The
SNR of 20 was obtained with an integration time per pixel of 10~ms
and averaging of 20 scans. The laser intensity in the tip was
stabilized to $0.3\%$ by monitoring a slight light leakage around a
fiber bend in the cryostat.

The transmission spectrum recorded on $PD_{21}$ suggests that, as
expected intuitively, the incident light is ``absorbed" by the
molecule on resonance. However, scattering
theory~\cite{Jackson-book,Loudon} tells us that the signal $I_{\rm
d}$ on the detector $PD_{21}$ is due to the interference of the
electric fields of the excitation light $E_{\rm e}$ and the
radiation emitted by the molecule $E_{\rm m}$. Using the operator
notation common in quantum optics~\cite{Loudon}, we can write
\begin{equation}
I_{\rm d}=\left\langle \widehat{\mathbf{E}}_{\rm e}^{-}\cdot\widehat{\mathbf{E}}_{\rm e}^{+}\right%
\rangle +\left\langle \widehat{\mathbf{E}}_{\rm
m}^{-}\cdot\widehat{\mathbf{E}}_{\rm m}^{+}\right\rangle +2
\real{\left\{ \left\langle \widehat{\mathbf{E}}_{\rm e}^{-}\cdot\widehat{\mathbf{E}}%
_{\rm m}^{+}\right\rangle \right\}.}
\end{equation} The first term represents the intensity $I_{\rm e}$ of the
excitation light on the detector. The part of the electric field
$E_{\rm m}$ on $PD_{21}$ is composed of two contributions, $E_{\rm
m}=E_{21}^{\rm c}+E_{21}^{\rm ic}$. The first component represents
the coherent elastic (Rayleigh) scattering process where the phase
of the incident laser beam is preserved. The second term takes
into account an incoherent part which is produced by the inelastic
interactions between the excitation light and the upper state of
the molecule~\cite{Loudon} as well as the emission in the phonon
wing of the $2 \rightarrow 1$ transition due to its phonon
coupling with the matrix~\cite{orrit:93}. So the second component
of Eq.~(1) represents the molecular emission intensity $I^{\rm
c}_{21}+I^{\rm ic}_{21}$. The last term of Eq.~(1) is responsible
for the interference between the excitation field and the coherent
part of the molecular emission $E_{21}^{\rm c}$ and is often
called the ``extinction" term~\cite{Jackson-book}. It follows that
if $E_{\rm m}$ is much weaker than $E_{\rm e}$, the extinction
term dominates the direct molecular emission intensity. Thus,
although the Stokes shifted fluorescence removes some energy out
of the $PD_{21}$ detection path, the dip in Fig.~2b is
predominantly due to interference.

We now develop a quantitative description of the transmission
signal on $PD_{21}$, keeping in mind the vectorial nature and
polarization properties of the light fields. The excitation
electric field along the unit vector $\mathbf{u}_{\rm d}$ at
position $\mathbf{r}$ of the detector can be written as
\begin{equation}
\left\langle \widehat{\mathbf{E}}_{\rm
e}^{+}(\mathbf{r})\right\rangle
=\left[\left\langle \widehat{\mathbf{E}}^{+}_{\rm e}(\mathbf{r}_{\rm m})\right\rangle \cdot%
\mathbf{u}_{\rm m}\right]~g~\mathbf{u }_{\rm d}.
\end{equation} Here $\mathbf{r}_{\rm m}$ is the position of the molecule, and $\mathbf{u}_{\rm
m}$ is a unit vector along its transition dipole moment. We have
introduced $g=\left\vert g\right\vert e^{i\phi _g}$ as a complex
modal factor that accounts for the amplitude, phase and
polarization of the tip emission at $\mathbf{u}_{\rm d}$, starting
from the projection of $\mathbf{E}_{\rm e}$ on $\mathbf{u}_{\rm
m}$. The electric field of the coherently scattered light is given
by~\cite{Loudon}
\begin{equation}
\left\langle \widehat{\mathbf{E}}_{21}^{c+}(\mathbf{r})
\right\rangle =(\sqrt{\alpha}d_{21})~\rho _{21}~f~\mathbf{u}_{\rm
d}\cdot
\end{equation} The quantity $d_{21}=\langle 2|\hat{D}|1\rangle$ is the matrix element of
the dipole operator. In a solid-state system, the intensity of the
zero-phonon line is typically reduced by the Debye-Waller factor
$\alpha$ due to emission into the phonon wing~\cite{orrit:93}. The
dipole moment corresponding to the zero-phonon transition
therefore, is given by $\sqrt{\alpha}d_{21}$. Our preliminary
measurements for DBATT in \emph{p}-terphenyl thin films let us
estimate $\alpha=0.25\pm 0.05$. Furthermore, adapting the standard
textbook treatment of resonance fluorescence to a three level
system, we find for the steady-state density matrix element $\rho
_{21}$,
\begin{equation}
\rho _{21}=\frac{\Omega \left( -\Delta +i\gamma /2\right)
}{2}\mathcal{L}(\nu)
\end{equation}
with
\begin{equation}
\mathcal{L}(\nu)=\frac{1}{\Delta ^{2}+\gamma ^{2}/4+\Omega
^{2}(\gamma /2\gamma _{0})K} \simeq \frac{1}{ \Delta ^{2}+\gamma
^{2}/4}
\end{equation} where $\Delta=\nu-\nu_{21}$ is the detuning between
the laser frequency $\nu$ and the zero-phonon frequency
$\nu_{21}$. The quantity $\Omega =(\sqrt{\alpha}d_{21})\left[
\left\langle \widehat{\mathbf{E}}^{+}_{\rm e}(\mathbf{r}_{\rm m})\right\rangle \cdot%
\mathbf{u}_{\rm m}\right]/h$ stands for the Rabi frequency. We
have verified that all our measurements are in the regime well
below saturation so that $\Omega \ll\gamma$. The parameter
$K=1+k_{23}/2k_{31}$ accounts for the competition of various decay
rates $k_{21}$, $k_{23}$, and $k_{31}$ (see Fig.~1b). In our case
$k_{31}\gg k_{23}$ so that $K\simeq 1$. These considerations
result in the last form of the expression in Eq.~(5). Finally, in
Eq.~(3) we have introduced $f=\left\vert f\right\vert e^{i\phi
_{f}}$ similar to $g$, as a complex modal factor that determines
the angular dependence, the phase and polarization of the
molecular field at the detector. However, contrary to $g$, we have
included some dimensional parameters in $f$ to simplify the
presentation of our formulae.

Using Eqs.~(1-5), we can now write
\begin{eqnarray}
I_{\rm d}&=&I_{\rm e}\left[1+\alpha~d_{21}^{4}~\frac{\gamma
}{\gamma _{0}}\left\vert \frac{f}{g} \right\vert
^{2}\mathcal{L}\left( \nu \right)\right. \nonumber\\ &-& \left.
2~\alpha~d^2_{21}\left\vert \frac{f}{g} \right\vert
\mathcal{L}\left( \nu \right) \left( \Delta \cos \psi +
\frac{\gamma }{2}\sin \psi\right)\right] \end{eqnarray} for the
forward direction. The phase angle $\psi=\phi_f-\phi_g$ denotes
the accumulated phase difference between the radiation of the
molecule and of the SNOM tip on the detector after propagation
through the whole optical system. By determining $I_{\rm e}$ from
the off-resonant part of the spectra and $\gamma$ from the
simultaneously recorded $I_{23}(\nu)$, we have fitted our
frequency scans using Eq.~(6), whereby we have used $\psi$ and the
multiplicative factor $d^2_{21}\left \vert f/g\right \vert$ as
free parameters. An example is shown by the red solid curve in
Fig.~2b where we obtained $\psi=\pi/2$. It is also possible to
vary $\psi$ to obtain dispersive shapes or even peaks. One way of
achieving this is to scan the tip in the \emph{x, y}, and \emph{z}
directions; our results on such studies will be published
separately. In what follows, we show how the observed signal could
be changed by selecting different polarizations in the detection
path. Furthermore, we define the visibility $V(\nu)=(I_{\rm
d}-I_{\rm e})/I_{\rm e}$ of the detected signal and will discuss
how the change in the ratio $\left\vert f/g \right\vert$ of the
modal factors influences the observed visibility.

\begin{figure}[b!]
\centering
\includegraphics[width=6.5 cm]{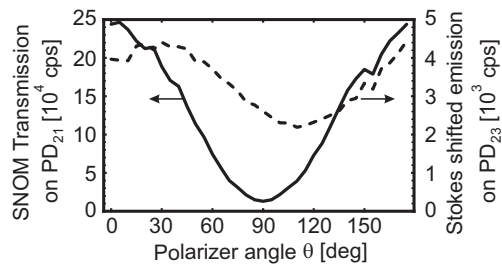}
\caption{Tip emission and Stoke shifted molecular fluorescence for
different orientations of a polarizer in the detection.}
\end{figure}

Figure 3 shows the polarization dependence of the SNOM emission
and the Stokes shifted fluorescence on the two detectors when a
linear polarizer was placed after the cryostat (see Fig.~1a). We
find that the main polarization axes of $E_{\rm e}$ and $E_{23}$
are offset by about $20^\circ$. The weak polarization extinction
ratio (2:1) of $I_{23}$ stems partly from the component of the
molecular dipole along the optical axis which gives rise to
radially polarized fields at the outer part of the detected
beam~\cite{Lieb:04}. Furthermore, the \emph{p}-terphenyl film and
dielectric mirrors introduce a non-negligible degree of
ellipticity to the polarization. Note that since the polarization
properties of the fields $E_{23}$ and $E^{\rm c}_{21}$ of the
molecule are both determined by the orientation of its dipole
moment, the dashed lines in Fig.~3 also provide information on the
polarization properties of $E^{\rm c}_{21}$.

In Fig.~4a we present $V(\nu)$ recorded on $PD_{21}$ for a series
of polarizer angles $ \theta$ whereby $\theta=0$ marks the highest
$I_{\rm e}$. Here we have adapted the integration times for each
$\theta$ to keep roughly the same SNR. As $\theta$ changes, the
spectrum evolves from an absorptive to a dispersive shape,
revealing a variation in $\psi=\phi_f-\phi_g$. The points in
Fig.~4b display an example for $\theta=75^\circ$. The rotation of
the polarizer results in the projection of $E_{\rm e}$ and $E^{\rm
c}_{21}$ onto different detection polarizations $\mathbf{u }_{\rm
d}$, which in turn implies variations of $g$ and $f$. In
particular, $\psi$ is changed if the fields possess some
ellipticity. Indeed, by extracting the degree of ellipticity from
the polarization data in Fig.~3, we have been able to fit the
recorded spectra in Fig.~4a simultaneously for all $\theta$ using
Eq.~(6).

As displayed in Fig.~4c, a remarkable situation results for the
cross-polarized detection ($\theta=90^\circ$) where the visibility
reaches $+10\%$ on resonance, yielding $\psi\simeq 3\pi/2$. This
surplus of light on resonance is a clear signature of interference
and redistribution of light. In fact, we have verified that the
sum of the spectra recorded at $\theta$ and $\theta+\pi/2$ remains
constant and absorptive for all $\theta$. The fact that $V$ is
larger here than in Fig.~2b although $I_{\rm d}$ is diminished to
$1/20$ of its original value, can be readily explained by Eq.~(6)
where reducing $g$ causes an increase in $V$. It is furthermore
interesting to note that if $g$ approaches zero, the second term
in Eq.~(6) dominates, giving access to the direct emission on the
$2 \rightarrow 1$ transition. The physical origin and the
optimization of $I_{\rm d}$ and $V$ therefore, depend on the ratio
$\left\vert f/g \right\vert$.

We now discuss various aspects of our results and their prospects
for future studies. The strong near-field coupling between the
laser field and the molecule makes it possible to detect the
coherent emission of weakly fluorescing emitters. The data in
Fig.~2 show that a very low coherent emission of $I^{\rm
c}_{21}\sim25$ counts per second (cps) could be detected at a
comfortably measurable contrast of $1\%$ with an incident power of
$I_{\rm e}=2.5\times 10^5$ cps. On the other hand, an ideal
two-level system would deliver even a larger effect than that
reported here. Since the extinction term in Eq.~(6) is
proportional to $\alpha/\gamma$ on resonance, $\gamma_0=8 \pm 1
$~MHz and $\gamma=35$~MHz (see Fig.~2) let us estimate that $V$
could have been $ \gamma/(\alpha\gamma_0)\simeq16$ times larger,
approaching $100\%$.

\begin{figure}[b!]
\centering
\includegraphics[width=7 cm]{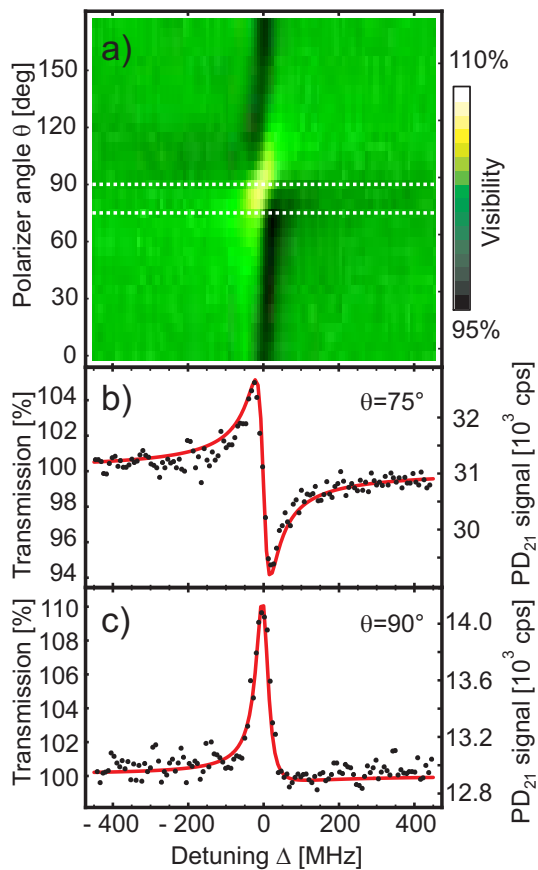}
\caption{(a) Frequency spectra recorded on $PD_{21}$ for 30
polarizer angles spanning $180^\circ$. The visibility is color
coded. (b, c) Cross sections from (a) for $\theta=75^\circ$ and
$\theta=90^\circ$ with fit functions according to Eq.~(6)}
\end{figure}

We emphasize that the visibility $V$ defined above is not a direct
measure of the absorption cross section that is conventionally
defined for a plane wave excitation. A rigorous treatment of the
absorption cross section for near-field excitation should take into
account the inhomogeneous distribution of the vectorial excitation
and molecular fields, as well as the tip and sample geometries, and
it remains a topic of future study. Here it suffices to point out
that the key advantage in near-field coupling is that the field
lines of the SNOM tip or a nanoscopic dipolar
antenna~\cite{Kuehn:06} are well matched to those of a dipolar
emitter, resulting in a better mode overlap than that achievable in
far-field optics~\cite{vanEnk:01}. In this work, we have considered
the forward propagation through an iris (see Fig.~1a). However, by
replacing the avalanche photodiode with a sensitive camera, it is
possible to map the complex mode overlap between $E_{\rm m}$ and
$E_{\rm e}$. We are currently pursuing such measurements.

The experiments presented here were performed on many samples,
molecules, and tips, and visibilities above $2\%$ were
consistently obtained. To the best of our knowledge, an extinction
of $6\%$ and a surplus signal of $10\%$ are the largest effects
yet reported for a single emitter acting directly on a light beam.
The interferometric nature of our detection scheme provides access
to the coherent optical phenomena in a single solid-state emitter.
Furthermore, the efficient near-field absorption spectroscopy
technique presented here can be used to study weakly fluorescing
nano-objects or single emitters that do not have large Stokes
shifts. Moreover, our work might find applications in single atom
detection on atom chips~\cite{Horak:03}. Finally, a direct and
strong coupling of light to a single emitter allows an efficient
manipulation of its phase and amplitude, which are considered to
be key elements for the execution of various quantum optical
tasks~\cite{Turchette:95}.

We thank M. Agio for theoretical and A. Renn, C. Hettich and S.
K\"{u}hn for experimental support. This work was financed by the
Schweizerische Nationalfond (SNF) and the ETH Zurich initiative
for Quantum Systems for Information Technology (QSIT).

\end{document}